\documentclass[12pt]{article}

\def\fnote#1#2{\begingroup\def\thefootnote{#1}\footnote{#2}\addtocounter{footnote}{-1}\endgroup}

\usepackage{amsmath}
\usepackage{amssymb}
\usepackage{makeidx}
\usepackage{epsf}
\usepackage{graphicx}        % standard LaTeX graphics tool
\makeindex

\def\inbar{\vrule height1.5ex width.4pt depth0pt}
\def\IB{\relax{\rm I\kern-.18em B}}
\def\IC{\relax\,\hbox{$\inbar\kern-.3em{\rm C}$}}
\def\ID{\relax{\rm I\kern-.18em D}}
\def\IE{\relax{\rm I\kern-.18em E}}
\def\IF{\relax{\rm I\kern-.18em F}}
\def\IG{\relax\,\hbox{$\inbar\kern-.3em{\rm G}$}}
\def\IH{\relax{\rm I\kern-.18em H}}
\def\II{\relax{\rm I\kern-.18em I}}
\def\IK{\relax{\rm I\kern-.18em K}}
\def\IL{\relax{\rm I\kern-.18em L}}
\def\IM{\relax{\rm I\kern-.18em M}}
\def\IN{\relax{\rm I\kern-.18em N}}
\def\IO{\relax\,\hbox{$\inbar\kern-.3em{\rm O}$}}
\def\IP{\relax{\rm I\kern-.18em P}}
\def\IQ{\relax\,\hbox{$\inbar\kern-.3em{\rm Q}$}}
\def\IR{\relax{\rm I\kern-.18em R}}
\def\IT{\relax{\rm I\kern-.18em T}}
\def\ZZ{\relax{\sf Z\kern-.4em Z}}

\def\nablaslash{\relax{\rm /\kern-.28em \nabla}}

\def\a{\alpha}         
\def\e{\epsilon} \def\G{\Gamma}

\def\cC{{\cal C}} \def\cD{{\cal D}}  
 \def\cH{{\cal H}}  
 \def\cL{{\cal L}}

\def\mathR{{\mathbb R}}

 \def\oD3{{\overline \rmD 3}}

   \def\phidot{{\dot{\phi}}}

\def\beq{\begin{equation}}
\def\eeq{\end{equation}}
\def\bea{\begin{eqnarray}}
\def\eea{\end{eqnarray}}

\def\lleq#1{\label{#1}\eeq}

\def\notin{\ \hbox{{$\in$}\kern-.51em\hbox{/}}}
\def\notsubset{\ \hbox{{$\subset$}\kern-.63em\hbox{/}}}

\def\del{\partial}

  \def\E1Fq{E_1/\IF_q}

\def\rmD{{\rm D}}

           \def\rmem{{\rm em}}

 \def\rmD{{\rm D}}

\def\rmkin{{\rm kin}}

   \def\rmsr{{\rm sr}}

      \def\rmIm{{\rm Im}}

        \def\rmPl{{\rm Pl}}

\def\notdiv{{\relax{~|\kern-.35em /~}}}

\def\boxit#1{
\vbox{\hrule height1pt\hbox{\vrule width1pt\kern0.3cm
\vbox{\kern0.3cm\hbox{$\displaystyle#1$}\kern0.3cm}\kern0.3cm\vrule
width1pt}\hrule height1pt}}

\begin{document}

\parindent=0pt

%Wk 20

 \phantom{whatever \hfill  \today~}

\vskip 1.2truein 

\centerline{\bf {\Large The Swampland Spectrum Conjecture in Inflation}}

\vskip .4truein

\centerline{\sc Rolf Schimmrigk\fnote{1}{rschimmr@iusb.edu}}

\vskip .3truein

\centerline{Dept. of Physics}

\vskip .1truein

\centerline{Indiana University at South Bend}

\vskip .1truein

\centerline{1700 Mishawaka Ave., South Bend, IN 46634}

\vskip 1truein
\baselineskip=18pt

\parskip=0pt
\centerline{\bf Abstract}
\begin{quote}
The quantum gravity conjectures that aim to separate the landscape from the swampland among the low energy theories 
 were originally formulated in the context of scalar field spaces spanned by moduli. Because these 
conjectures have implications for cosmology they have recently been considered in a more general context for 
 scalar field theories with potentials, in particular inflation. From an effective field theory perspective the presence 
 of a potential induces a natural metric that makes the distance measure $D_V$ along scalar field trajectories dependent
 on the potential. The present paper proposes a modified formulation in terms of $D_V$ 
 of  those conjectures that involve trajectory distances.
\end{quote}

\renewcommand\thepage{}
\newpage
\parindent=0pt

 \pagenumbering{arabic}

\baselineskip=15pt
\parskip=0.pt

\tableofcontents

\vskip .5truein

\parskip=0.1truein
\baselineskip=19.5pt 

\section{Introduction}

Recent discussions in the literature have attempted to address  the possible implications of the conjectures that 
aim to separate low energy theories that admit UV completions from those that do not. The former are said to belong 
to the quantum gravity landscape, the latter to the swampland \cite{v05, ov06,  b17etal, o18etal, oo18etal}. 
Two  of these conjectures are concerned with the distance that scalar fields traverse in some field space $X$. 
The first of these is the infinite diameter conjecture \cite{v05, dl05}, which posits that the target space should 
admit trajectories of infinite length. Building on this is a second conjecture that has been the focus of much attention 
and is in the earlier literature referred to as the swampland conjecture, and more recently, after the advent of 
\cite{o18etal} as  the swampland distance conjecture. This conjecture is concerned with the mass spectrum of the theory 
as the fields evolve and will here be referred to simply as the 
 spectrum conjecture to distinguish it from the infinite diameter conjecture on which it builds. 
 
 A precise formulation of the spectrum conjecture has not stabilized yet, but very roughly it states that for a given low-energy field theory 
 the Lagrangian remains valid only if the evolution traverses a distance smaller than the Planck scale. The picture assumed here 
 is that the mass spectrum changes as the scalar field evolves from a point $p_0$ in the field space $X$ to a point $p$ at 
 distance $\cD(p_0,p)$ from $p_0$ because light particles appear whose masses scale with the distance. In the original 
 asymptotic formulation for a single scalar field with a flat target space the distance dependence is taken to be
 \beq
  m(p) ~\cong ~ M_\rmPl e^{-\a \cD(p_0,p)/M_\rmPl},
 \eeq
 where $\a$ is an undetermined positive parameter that at present has to be estimated in a model dependent way and
 $M_\rmPl$ is the reduced Planck scale. 
 The main issue that is unresolved in this formulation is at what point the exponential behavior should be expected to set it.
 In the simplest case the spectrum conjecture is formulated as an asymptotic statement for a single scalar field $\phi$,
 i.e. the functional form of the masses is assumed to be valid in the limit in which the flat target space 
 distance $\cD = \Delta \phi$ diverges,  
 and in this limit an infinite tower of light particles is envisioned to appear \cite{ov06}.  An attempt to 
encode these local structure of the moduli space has been made in \cite{kp16}, where a further unknown 
function is introduced in addition to the exponential factor in order to parametrize the current ignorance of the local effects. 

 Originally the formulation of the landscape vs. swampland  discussion was framed in the context of string theory, 
 where the main focus is on scalar fields $\phi^I$ that are moduli 
          (see e.g.  \cite{kp16,  bvw17, p17, h17etal, g18etal, b18etal, bl18, landete-shiu18, l18etal-a, l18etal-b} 
           for recent work and additional references).
 In this context the distance $D(p_0,p)$ is  measured with the metric on the moduli space, given by the
   metric $G_{IJ}$ of the kinetic term
 \beq
  \cL_\rmkin ~=~ - \frac{1}{2} G_{IJ}(\phi^K) g^{\mu\nu} \del_\mu \phi^I \del_\nu \phi^J.
 \eeq
Recently however  both the infinite diameter conjecture and the associated spectrum conjecture have been considered
 in the context of more general field theories, in particular in the framework of inflation, where one of the issues that has
 been discussed extensively is concerned with large field inflation. The spectrum conjecture has been addressed 
 both in the singlefield framework  \cite{h07, a18etal, kr18, d18etal, mt18, k18etal,   bh18, d18, kt18, m18etal, a18}
 and in the multifield generalization \cite{n08, b15etal, ap18, b18}.  Related work includes 
   \cite{gk18, lh18etal, l18etal,   d18b, gkz18, yg18, clm18, k18, h19etal, abl19, a19etal, m19etal, w19etal}.
   In this more general case  the existence of a potential 
changes the nature of the trajectory because the inflaton rolls down the potential surface, which in an $n$-component 
scalar field theory is a hypersurface embedded in an $(n+1)$-dimensional space. In this framework the potential 
induces a natural metric on the potential surface that leads to a modified distance
 formula, involving both the field space metric $G_{IJ}$ as well as a term that is induced by the gradients 
 of the potential $V(\phi^I)$. The purpose of this note is to propose a modified swampland distance conjecture 
 that takes the potential $V$ of the model into account via a modified distance $D_V$ 
  and to discuss its implications in the context of the quantum gravity vs. 
 swampland conjectures.

\section{Field space trajectories}

In curved multifield theories the target space is a configuration space $X$ that is equipped with a 
Riemannian metric $G_{IJ}$
and the fields are constrained by a potential $V(\phi^I)$, where the number of fields $I,J =1,...,n$ is arbitrary. 
 The Klein-Gordon 
evolution of the background field $\phi^I(t)$  
  \beq
   D_t\phidot^I ~+~ 3H\phidot^I ~+~ G^{IJ}V_{,J} ~=~ 0
  \eeq
  then leads to trajectories in the field space $X$. Here $D_t$ is the covariant derivative, defined on a vector 
  field $W^I$ as
  $D_tW^I = \del_t W^I + \G^I_{JK} \phidot^J W^K$, where the connection is assumed to be of Levi-Civita type, 
  and the connection
  coefficients   $\G^I_{JK}$ are the associated Christoffel symbols.  
  The distances in the target space $X$ are given by the standard length formula
 \beq
  \cD ~=~ \int_{t_0}^{t_e} dt ~\sqrt{G_{IJ}(\phi^K) \phidot^I \phidot^J}.
 \eeq

 A concrete illustration of this picture can for example be given within the class of automorphic inflation \cite{rs14,rs15}, 
 in which case the target space is obtained via group quotients that endow $X$ with a discrete symmetry that 
 is constrained by the fact that it contains the  shift symmetry. In the special case of modular inflation the field space 
 is the complex upper halfplane $\cH$,
 spanned by an inflaton doublet $(\phi^1, \phi^2)$, on which the geometry is determined by the Poincar\'e metric 
 \beq
  G_{IJ} ~=~ \frac{1}{(\rmIm ~\tau)^2} \delta_{IJ}
 \eeq
 where the dimensionless variables  $\tau^I = \phi^I/\mu$ introduce a mass scale $\mu$ that is constrained by 
  CMB data. Potentials in this framework are given by a modular invariant function, for example the 
  $j$-inflation potential  $V =\Lambda^4 |j|^2$, where $j(\phi^I)$ is the absolute modular invariant \cite{rs14, rs16, rs17}.  
  Modular invariance  suggests to combine the inflaton multiplet into the dimensionless complex variable $\tau$.
  Generalizations to modular inflation at higher level are introduced in \cite{ls19}.

 Inflationary trajectories in the modular inflation target space $X=\cH$ are shown in Fig. 1 
  for the potential of $j$-inflation. (Graphs for trajectories of higher level inflation can be found in ref. \cite{ls19}.)
  Their length is obtained by computing 
   \beq
   \cD ~=~ \mu \int dt ~\frac{1}{\rmIm ~\tau} \sqrt{\left(\frac{d\tau^1}{dt}\right)^2 + \left(\frac{d\tau^2}{dt}\right)^2}
  \eeq
  along the inflaton path. These paths start in a region of the inflaton space where the slow-roll parameters are small and 
  they continue until the parameter $\e$ approaches unity.

 \begin{center}
  \includegraphics[scale=0.4]{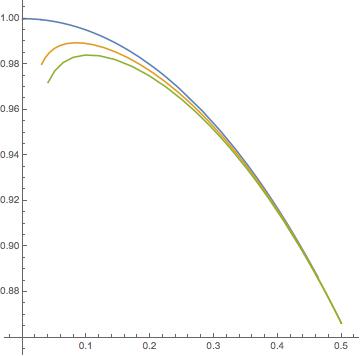}
 
 \begin{quote}
 {\bf Fig. 1.}~{\it Phenomenologically consistent inflaton trajectories $(\tau^1, \tau^2)$ 
      on the target space $X=\cH$ of the upper halfplane for $j$-inflation. More details can be found 
      in \cite{rs16}.}
\end{quote}
 \end{center} 
 
\section{Trajectory distances with potentials}

In the case of scalar field theories with a potential $V(\phi^I)$ the trajectories in the target space  $X$ 
 are not the trajectories that encode the physics of the theory because they represent a lower-dimensional projection 
 of the path traversed by the inflaton on the potential surface associated to $V$. Thinking of $V$ as a function that is 
 defined by the field target space $X$ and takes values in the set of real numbers $\mathR$ shows that the presence of the 
 potential adds another dimension, leading  to the space $X\times \mathR$ as the relevant configuration space. The potential 
 surface is defined as a hypersurface embedded in this space. For the case of flat target spaces considered in 
 many papers on multifield inflation with $n$ fields this product space simply corresponds to the $n$-dimensional euclidean 
 space $\mathR^{n+1}$. In multifield theories with curved targets $X$ the embedding space usually has a 
different topology, depending on the symmetries of the models. 

The presence of a potential implies that the physical length of the inflaton paths is naturally measured along the 
inflaton trjajectores  on the potential surface because it is this surface that determines the physical model. 
 Depending on the precise structure of the potential the projection of the 
hypersurface trajectories onto the target space trajectories can provide a reasonable approximation of the actual paths on 
the potential surface
 when the gradients of the potential are small over the whole time interval. If on the other hand the variation of the 
 potential normal to the target space is significant then the hypersurface trajectory lengths can be quite different from the 
 distances traversed by the scalar field multiplet in the target space $X$.  
 Current CMB observations \cite{a18etal-a, a18etal-b}
  are consistent with slow-roll  inflation, in which case at the beginning 
 of inflation the slow-roll parameters 
 \beq
  \e_I ~=~ M_\rmPl \frac{V_{,I}}{V}
  \eeq
are small, hence the gradients of the potential are small.  If however inflation ends when the slow-roll 
parameter $\e$ approaches 
unity, as it does in many classes of theories, these parameters are no longer small.

An example of a potential hypersurface is shown in Fig. 2 for a model in the class of modular inflation theories
 considered in \cite{rs14,rs16,rs17}. The target space is again the complex upper halfplane $X=\cH$  with the above 
projected 2D trajectories in Fig. 1, but now the inflaton components are shown  as they evolve on the potential
 hypersurface. Fig. 2 illustrates that the 
 length of the inflaton trajectory can be quite different from the length of 
its projection in the target space, depending on the structure of the potential. It exemplifies that the gradients $V_{,I}$ 
of the potential can vary considerably along the inflaton trajectory in the space $X\times \mathR$.  
(In ref. \cite{ls19} trajectories are considered for a modular inflation model at a congurence subgroup.) 
Given that the physical model is characterized by both the field space metric $G_{IJ}(\phi^K)$ and the potential 
$V(\phi^I)$, it is natural to take the effect of the potential into account and consider 
the distances on the potential hypersurface as the fundamental quantities. The geometry of such hypersurfaces 
is classical and can in the simplest case of surfaces in flat three-dimensional space 
 be traced back to the beginnings of differential geometry with Gau\ss.

 \begin{center}
  \includegraphics[scale=0.6]{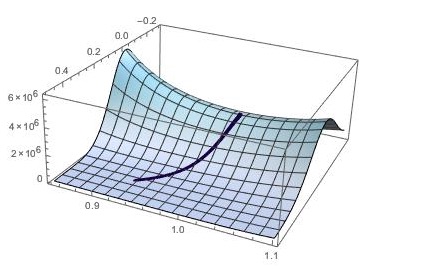} 
 \end{center}
 \begin{quote}
 {\bf Fig. 2.}~{\it  Trajectory with $N=60$ e-folds and {\sc Planck} compatible observables 
 on the potential surface of $j$-inflation with $V=\Lambda^4 |j|^2$.
  }
 \end{quote}

Given a multifield inflationary model with a target space metric $G_{IJ}$ and a defining potential $V(\phi^I)$
 as a function on the target space, a choice has to be made about how precisely to parametrize the 
resulting ambient space.  
A useful way to think about potentials in arbitrary multifield scalar theories  is by factoring out
 an overall scale $\Lambda$ and to 
 write the potential as a product of $\Lambda^4$ and a dimensionless function. 
 The target space $X$ is parametrized by the inflaton multiplet $\phi^I$ and the  trajectory surface $S^V$ associated 
 to the potential $V$  in the space $X\times \mathR$ is defined by the vector 
 \beq
  f^V(\phi^I) ~=~ (\phi^I, ~V(\phi^I)/\Lambda^3).
 \eeq
   This surface leads to a metric $G_{IJ}^\rmem$  that is induced by the metric of the embedding space $
   X\times \mathR$ as
   \beq
    G_{IJ}^\rmem ~=~ \langle f_{,I}^V, f_{,J}^V\rangle,
   \eeq
   where the derivatives are  defined with respect to the inflaton components $\phi^I$.
    The inner product depends on the structure of the field theory considered. If the metric $G_{IJ}$ is flat then this product 
   is just the euclidean product, while in the curved case it  is no longer flat but induces the 
   embedding space metric
     \beq
   G_{IJ}^\rmem ~=~ G_{IJ} ~+~ \frac{1}{\Lambda^6} V_{,I}V_{,J}.
  \eeq
  
 This leads to a modified distance formula that is induced by both the target space metric and the potential as
  \beq
   \cD_V(t_i, t_e) ~=~ \int_{t_i}^{t_e} dt ~\sqrt{G_{IJ} \phidot^I \phidot^J ~+~ \frac{1}{\Lambda^6} (\phidot^I V_{,I})^2}.
  \lleq{new-distance}
  This formula has an intuitively natural structure in that the larger the gradient of the potential the more the target space 
  distance $\cD$ will be different from $\cD_V$. This is precisely as expected from our conceptual discussion above.
  
  In the context of multifield slow-roll inflation the distance $\cD_V$ can be expressed purely as a function of the potential 
  and hence is amenable to an analytic treatment.  It is convenient to write the slow-roll approximation
   $\cD_V^\rmsr$ of $\cD_V$ in  terms of the slow-roll parameters $\e_I$ introduced above, which leads to
   \beq
  \cD_V^\rmsr ~=~ \int_{t_i}^{t_e} dt ~\sqrt{ \frac{V}{3} G^{IJ}\e_I\e_J  \left(1
    ~+~ \frac{1}{\Lambda^6} \frac{V^2}{M_\rmPl^2} (G^{IJ}\e_I \e_J) \right)}~.
  \lleq{new-distance-sr}

In the above discussion the simplest possible choice was adopted for the embedding surface in the sense that the 
metric on $X\times \mathR$ was taken to be the metric on $X$ and the flat metric on $\mathR$. In principle it is of course
possible to consider a warped metric on the reals and furthermore to introduce off-diagonal components of the metric, 
should the physical situation warrant this.

  \section{The modified spectrum conjecture of quantum gravity}
  
  The discussion of the previous section suggests to consider the distance $\cD_V$ of eq. (\ref{new-distance}) in the 
  context of those conjectures that are affected by the distance measurements, most immediately the infinite diameter 
  conjecture \cite{v05, dl05, h17etal, g18etal,  h18etal} and the 
  spectrum conjecture \cite{ov06, g18etal,  b18etal, bl18}.
   A theory that is in compliance with the infinite diameter conjecture 
  relative to the target space metric automatically satisfies the conjecture relative to $\cD_V$ because the additional 
  term in $\cD_V$ is positive. For the spectrum conjecture on the other hand the additional term matters because 
  of its more discriminating quantitative nature. 
  
  The present proposal for a refined formulation of  the asymptotic statement is to restate the conjecture
  as the expectation that in terms of the distance measure $D_V$ introduced above 
  the mass spectrum  introduces new light states according to the relation 
  \beq
   m ~\cong ~ M_\rmPl \exp\left(- \frac{\a}{M_\rmPl} \cD_V(\phi_i,\phi_e) \right).
  \eeq
  In the slow-roll approximation this factors for a small potential gradient term as 
  \beq
  e^{-\a \cD_V/M_\rmPl} ~\cong ~ \cC_V e^{- \a \cD/M_\rmPl},
  \eeq
  where $\cC_V$ encodes the potential gradient term. 
  The numerical factor $\a$ is not known, but has been conjectured to be of order one, thereby making the assumption 
  that no factors like   $4\pi^2$ will appear \cite{kp16}. 
 While the mass scale considered in the spectrum conjecture is canonically chosen to be the Planck mass $M_\rmPl$, 
  the natural scale in quantum gravity, one might expect that there is a local mass scale $m(\phi_0)$ that enters, 
  as considered in \cite{kp16}, where the local structure of the moduli space is furthermore parametrized in the 
  context of a single scalar field by an additional   factor $\G(\phi_0, \Delta \phi)$.
  In the present context this local factor takes the form $\G(\phi_0^I, \cD_V(\phi^I))$ with $\cD_V(\phi^I)$ as in 
  eq. (\ref{new-distance}).
  
  Current CMB observations \cite{a18etal-a, a18etal-b} are consistent with slow-roll inflation, for which the relative gradients
   $\e_I = M_\rmPl V_{,I}/V$ are small. Furthermore, these parameters should vary slowly so as to achieve the standard
    range of e-folds, usually required to be bounded by  $N_e\in [50,~70]$.   In many models inflation ends eventually
  and the end is reached as the slow-roll parameter $\e$ approaches 
  unity. This means that the slow-roll approximation will fail toward the end of the trajectory and for those regions 
  the potential term in $\cD_V$ will be significant.
  
  The spectrum conjecture can be related to the weak gravity conjecture \cite{a06etal} under the assumption  that the  
  gauge coupling varies with the scalar field evolution \cite{kp16}.  This relation builds on the general expectation that in a 
  quantum  gravity context continuous  coupling parameters are determined by vevs of scalar fields, an assumption 
  that has been encoded as conjecture zero in the swampland literature \cite{ov06}. In its simplest and 
    most robust form the WGC states that 
   in any consistent theory of quantum gravity with a  U(1) gauge group there exists a particle whose mass is bounded by 
    \beq
     m ~\leq ~ qgM_\rmPl,
    \eeq
   where $g$ is a dimensionless measure of the U(1) coupling parameter and $q$ is the relative charge.
    If the coupling parameter is a function of some scalar
   fields $\phi^I$ then the evolution of these fields of course changes the mass of at least this particular particle. 
   While in our universe this relation holds with a very wide margin, indicating that a further ingredient of the gravity story 
   is missing, this conjecture has been explored extensively for its possible implications for large field inflation.
  In a general formulation of the weak gravity conjecture the distance enters and therefore the consideration of 
  $\cD_V$ also has implications for the conceptual development of the weak gravity conjecture. Finally, the spectrum 
  conjecture has very recently also been related to the de Sitter gradient conjecture in ref. \cite{oo18etal}.
  
\vskip .2truein

{\bf Acknowledgement.} \\
It is a pleasure to thank Monika Lynker for discussions.
This work was supported in part by a Faculty Research Grant at Indiana University South Bend.

\end{document}